\documentclass[aps,prl,twocolumn,showpacs,groupedaddress]{revtex4}  % for review and submission 
\usepackage{graphicx}  % needed for figures
\usepackage{dcolumn}   % needed for some tables
\usepackage{amssymb,amsmath}   % for math

\begin{document}

% the following line is for submission, including submission to the arXiv!!
%\hspace{5.2in} \mbox{Darmstadt/DESY/U-Tokyo}

\title{Quantum theory of fermion production after inflation}
\author{J{\"u}rgen Berges}
\author{Daniil Gelfand}
\author{Jens Pruschke}
\affiliation{Institute for Nuclear Physics, Darmstadt University of Technology, Schlossgartenstrasse\ 9, 64285 Darmstadt, Germany}

\begin{abstract}
We show that quantum effects dramatically enhance the production of  
fermions following preheating after inflation in the early Universe in  
the presence of high excitations of bosonic quanta. As a consequence fermions rapidly approach a quasistationary distribution with a thermal occupancy in the infrared, while the inflaton enters a turbulent scaling regime. The failure of standard semiclassical descriptions based on the Dirac equation with a homogeneous background field is caused by nonperturbatively high boson occupation numbers. During preheating the inflaton occupation number increases, thus leading to a dynamical mechanism for the enhanced production of fermions from the rescattering of the inflaton quanta. We comment on related phenomena in heavy-ion collisions for the production of quark matter fields from highly occupied gauge bosons. 
\end{abstract}

\maketitle

The precise understanding of phenomena out of equilibrium plays
a crucial role for our knowledge about the primordial Universe. Important examples
are the generation of density fluctuations, nucleosynthesis, or baryogenesis, with the latter being responsible for our own existence.

Most dramatic processes such as far-from-equilibrium particle production at the end of 
inflation pose a particular theoretical challenge~\cite{Traschen:1990sw}. The corresponding nonperturbative phenomenon of preheating is conventionally described using classical approximations for the bosonic inflaton~\cite{Khlebnikov:1996mc}. Their validity for  macroscopic occupation numbers has been verified explicitly in quantum field theory~\cite{Berges:2002cz}. Much less is known about fermion dynamics in the nonperturbative regime of high Bose occupation numbers.
Since identical fermions cannot occupy the same state, their quantum nature is highly relevant and a consistent quantum theory of fermion production after inflation is of crucial importance. Related questions arise in the context of heavy-ion collision experiments, where the production of quark matter fields from highly occupied gauge bosons can occur far from equilibrium~\cite{RHIC}. 

So far, preheating dynamics with fermions has been mainly investigated based on semiclassical descriptions using the Dirac equation with coupling to a homogeneous inflaton field~\cite{Baacke:1998di}. Also backreaction of fermions onto inflaton dynamics has been included. A similar semiclassical prescription has been used to compute quark production from classical gluons in the context of heavy-ion collisions~\cite{Gelis:2005pb}. Further inclusion of quantum corrections is complicated by a secular perturbative time evolution which becomes rapidly invalid. Suitable approximations may be based on the two-particle irreducible (2PI) effective action, which are known to describe the approach to thermal equilibrium~\cite{Berges:2002wr}. Simplified discussions with these techniques, which evaluate the dynamics for vanishing inflaton field amplitude, exhibit new nonlinear phenomena such as instability-induced fermion production \cite{Berges:2009bx}. 

In this work we present a quantum theory of fermion production following preheating after inflation. We consistently include quantum corrections to next-to-leading order (NLO) in the Yukawa coupling
between the inflaton field and massless fermions. Even for weak couplings this turns out to change semiclassical or leading-order (LO) results so dramatically that we consider a complementary nonperturbative method for comparison. It is based on lattice simulations following the techniques of Ref.~\cite{Aarts:1998td}. This treats the fermions exactly, but the inflaton dynamics becomes classical statistical. In $3+1$ dimensions this is computationally expensive and becomes feasible with the implementation of "low-cost" fermion algorithms~\cite{Borsanyi:2008eu}. Remarkably, higher-order corrections turn out to leave the NLO results practically unchanged for the considered range of weak couplings.  

In inflationary cosmology, the Universe at early times expands quasiexponentially
in a vacuumlike state. During this stage of inflation,
all energy is contained in a slowly evolving inflaton field. Eventually the
inflaton has to transfer its energy to particles, thereby
starting the thermal history of the hot Friedmann universe. If the creation of particles is sufficiently slow (for instance, if the inflaton is coupled only gravitationally to the matter fields) this transfer may be described by perturbative decay. However, for a wide range of couplings the particle production is governed by a nonequilibrium instability~\cite{Traschen:1990sw,Khlebnikov:1996mc,Berges:2002cz}. These instabilities are known to lead to exponential growth of inflaton occupation numbers in long wavelength modes on time scales much shorter than the asymptotic thermal equilibration time. This is followed by a turbulent phase with different universal scaling regimes for nonperturbative long wavelength modes~\cite{Berges:2008wm} and perturbative higher momenta~\cite{Micha:2002ey}. 
Despite the differences between early-universe dynamics and the "little bang"
generated in a heavy-ion collision, there are remarkable parallels concerning the role of instabilities and turbulence for the process of thermalization~\cite{Berges:2008pc}. 

To reveal the production of matter fields from inflaton decay requires the inclusion of fermions. We consider a quantum field theory with interaction of the Yukawa type 
\begin{equation}
- \frac{g}{N_f} \bar{\psi}_i \left( (1-\gamma^5) \Phi^\dagger_{ij} + (1+\gamma^5) \Phi_{ij} \right) \psi_j \, ,
\label{eq:interaction}
\end{equation}
for $i,j = 1,\ldots,N_f$ massless Dirac fermions, where summation over repeated indices is implied. (We use natural units with $\hbar = c = k_B = 1$.) The interaction is constructed to couple the left- and right-handed fermion field components $\psi_L = (1-\gamma^5)\psi/2$, $\psi_R = (1+\gamma^5)\psi/2$
for a nonvanishing expectation value 
\begin{equation}
\phi(t) = \langle {\mathrm{Tr}}\, \Phi(t,{\bf{x}}) \rangle
\end{equation}
of the inflaton field. This describes the generation of an effective fermion mass by spontaneous symmetry breaking. Together with the scattering or decay processes implied by (\ref{eq:interaction}), the interaction contains the major building blocks required to discuss fermion production. 

To be specific we consider two flavors, $N_f =2$, with symmetry group $SU_L(2) \times SU_R(2) \sim O(N_s=4)$, where $\Phi =\left(\sigma + i \vec{\tau} \vec{\pi} \right)/2$ with Pauli matrices $\vec{\tau}$ and an inflaton potential  
$V = m^2 (\sigma^2 + \vec{\pi}^2)/2 + \lambda\left(\sigma^2 + \vec{\pi}^2\right)^2/4!N_s$
characteristic for a linear sigma model. For our purposes it will be sufficient to consider dynamics without expansion. In chaotic inflationary models, after the end of inflation the inflaton field coherently oscillates around the minimum of its potential with a very large amplitude. The coupling of the field to its own quantum fluctuations leads to the phenomenon of parametric resonance~\cite{Traschen:1990sw}. For realistic scenarios the inflaton mass is required to be about 6 orders of magnitude smaller than the value of the inflaton field, and the inflaton quartic coupling is about $10^{-12}$. Since simulations are hard to perform for realistic hierarchies, we will consider numerical results where the mass is an order of magnitude smaller than the inflaton with a coupling $\lambda = 0.1$. The realistic case can then be inferred from our analytic results for time scales and production rates. Though all results will be for parametric resonance in chaotic inflation scenarios, we concentrate on characteristic properties which are independent of the underlying instability type and details of the model.   

Fermion production can be computed from 
\begin{equation}
F(x,y) = \frac{1}{2} \left\langle [ \psi(x), \bar{\psi}(y) ] \right\rangle , \,
\rho(x,y) = i \, \left\langle \{ \psi(x), \bar{\psi}(y) \} \right\rangle .
\label{eq:Fdef}
\end{equation}
Here $F(x,y)$ denotes the commutator or statistical two-point function, and $\rho(x,y)$ is the anticommutator of two fermion fields or the spectral function. In thermal equilibrium these would be translation invariant and related by a fluctuation-dissipation relation
in Fourier space: $F_{\rm eq}(\omega,{\bf p}) = -i [1/2 - n_{\rm FD}(\omega)] \rho_{\rm eq}(\omega,{\bf p})$ with the Fermi-Dirac distribution $n_{\rm FD}(\omega) = 1/[\exp(\omega/T)+1]$ for zero net fermion charge~\cite{Berges:2002wr}. The question of fermion production out of equilibrium may be discussed in terms of a time-dependent occupation number $n_\psi(t,{\bf p})$, which starts out at $n_\psi(t=0,{\bf p}) \simeq 0$ after inflation and approaches a constant distribution in equilibrium. Calling $M_\psi(t) = g \phi(t)/N_f$ we consider a standard definition~\cite{Baacke:1998di}
\begin{equation}
n_\psi(t,{\bf p}) = \frac{1}{2} - \frac{|{\bf p}| F_V(t,t;{\bf p}) + M_\psi(t) F_S(t,t;{\bf p})}{\sqrt{{\bf p}^2+M_\psi^2(t)}} \, ,
\end{equation}
where $F_S = \mathrm{tr}(F/4)$, $F_V = \mathrm{tr}( \vec{p} \cdot \vec{\gamma}\, F/4)$
with the trace acting in Dirac space. These are the nonvanishing components of $F$ at initial time consistent with symmetries.

\begin{figure}[t]
\includegraphics[scale=0.7,bb=50 50 410 302]{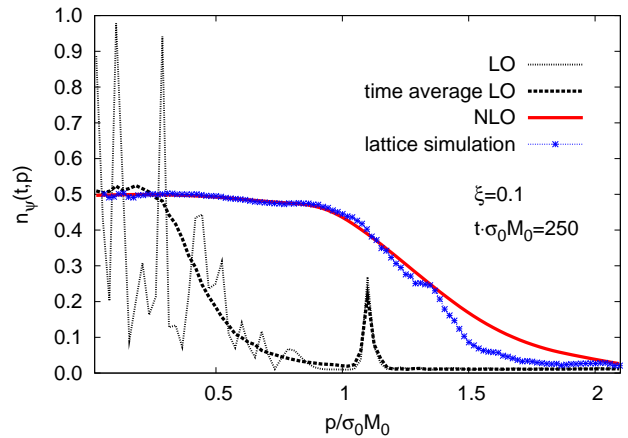}
\caption{Time-dependent fermion occupation number.}
\label{fig:nferm}
\end{figure}
The standard semiclassical (LO) evolution equation is
\begin{equation}
\left[ i \gamma^\mu \partial_\mu - \frac{g}{N_f} \phi(t) \right] F(x,y) \, = \, 0 \, .
\label{eq:FLO}
\end{equation}
Figure~\ref{fig:nferm} shows the corresponding LO result (dotted line) for the fermion occupation number for $\xi \equiv g^2/\lambda = 0.1$ at fixed time $t = 250/\sigma_0 M_0$ in units of the rescaled initial inflaton field $\phi_0 = \phi(t=0)$, i.e.\ $\sigma_0 M_0 = \phi_0/\sqrt{6 N_s/\lambda}$ with inflaton mass $M_0$. Since in this approximation the result is strongly oscillatory, in addition a time average (dashed line) over a period of about $20/\sigma_0 M_0$ is displayed. These characteristic LO properties of fermion occupation numbers, which vary periodically in time governed by resonant excitations, are discussed in Ref.~\cite{Baacke:1998di}.

In order to see whether the inclusion of quantum corrections changes these results, we first employ 2PI effective action techniques~\cite{Berges:2001fi,Berges:2002wr}. The fermion contribution in the coupling expansion of the 2PI effective action at NLO in $g$ reads diagrammatically:
\begin{eqnarray}
\includegraphics[scale=0.35]{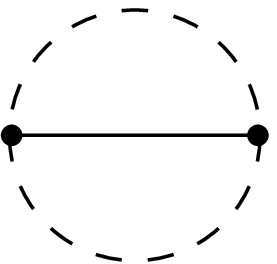}
\label{eq:fermapprox}
\end{eqnarray}
\noindent
Here the dashed lines denote self-consistently dressed fermion propagators and the solid line represents the inflaton propagator. For both LO and NLO results the same truncation for the bosonic sector is used, which includes all leading and subleading corrections in a $1/N_s$ expansion of the 2PI effective action:
\begin{eqnarray}
\includegraphics[scale=0.28]{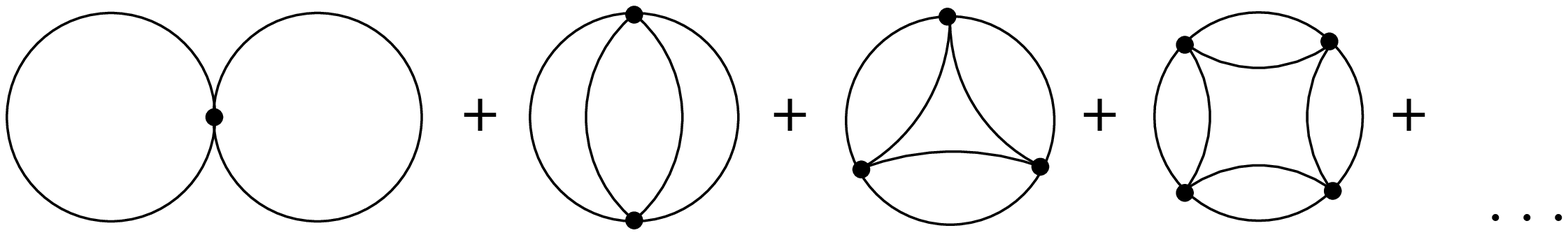}
\nonumber\\
\includegraphics[scale=0.28]{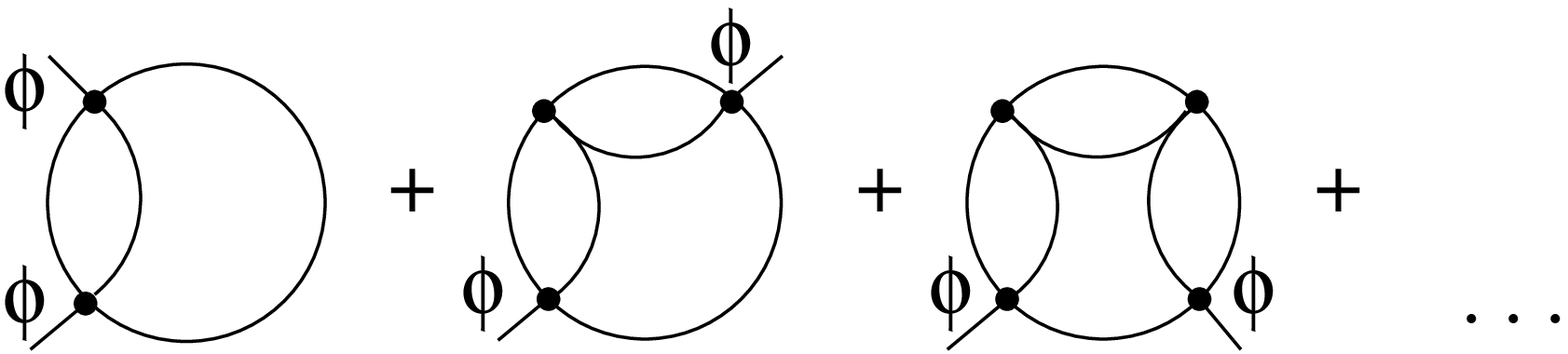}
\label{eq:bosapprox}
\end{eqnarray} 

\begin{figure}[t]
 \includegraphics[scale=0.7]{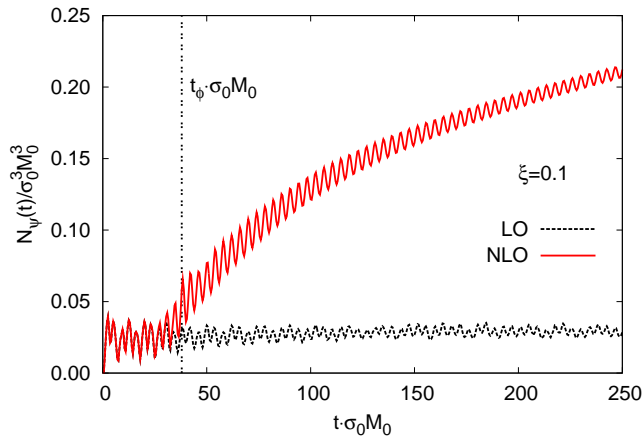}
\caption{Total number density of fermions.}
\label{fig:totalNferm}
\end{figure}

\begin{figure}[t]
\includegraphics[scale=0.7]{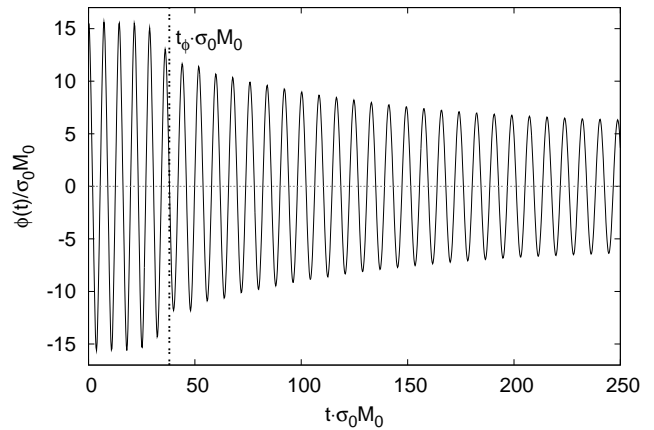}
\caption{Inflaton field amplitude.}
 \label{fig:phi}
\end{figure}

Figure~\ref{fig:nferm} shows that the NLO quantum correction (solid curve) changes the LO result for the distribution dramatically. To see how this difference builds up, we consider in Figure~\ref{fig:totalNferm} the total fermion number density $N_\psi(t) = 8 \int {\rm d}^3 p\, n_\psi(t,{\bf p})/(2\pi)^3$ as a function of time. Both LO (dashed) and NLO (solid) results agree rather well at early times at which parametric resonance leads to an exponential increase of the inflaton two-point function, $F_\phi (t,t;{\bf p}_0) \sim A_0 e^{2 \gamma_0 t}$. Here $\gamma_0$ denotes the growth rate of the maximally amplified momentum mode ${\bf p}_0$~\cite{Traschen:1990sw,Khlebnikov:1996mc,Berges:2002cz}. Significant deviations between the LO and NLO fermion production occur at the time $t_\phi$ where the inflaton field amplitude $\phi(t)$ starts to decrease substantially, as can be seen from Figure~\ref{fig:phi}. At this time also the $g \phi(t)$ contribution to the LO equation of motion (\ref{eq:FLO}) is diminished such that the field independent parts at NLO become important.

Given the somewhat unrealistic parameters for the numerical results, it is important to discuss what remains for realistic scenarios. For this we estimate characteristic time scales and production rates analytically.
Since fermions are not yet copiously produced at $t_\phi$, this time can be obtained from the bosonic sector along the lines of Ref.~\cite{Berges:2002cz}. For realistic hierarchies this yields
\begin{equation}
t_\phi \simeq \frac{1}{\gamma_0}\ln\left(\frac{\phi_0}{\sigma_0 M_0}\right) \, .  
\label{eq:tphi_PR}
\end{equation}  
For physically relevant $\phi_0/M_0 \sim 10^6$ and $\sigma_0 \sim O(1)$ this time is indeed of the same order as for the parameters in the numerics above. Apart from the fact that $t_\phi$ depends only logarithmically on the mass to amplitude ratio, it is well known that for the employed parametric dependence of the initial field $\phi_0 \sim 1/\sqrt{\lambda}$ the coupling can be scaled out of the equation of motion for the rescaled inflaton. It remains to estimate the enhanced fermion production rate after $t_\phi$. The amplification at this time is caused by the nonperturbative dependence of $F_\phi$ on the inverse of the coupling. 
Since the loop correction (\ref{eq:fermapprox}) is $\sim g^2$ and contains a boson line $\sim F_\phi$ of order $1/\lambda$, these together lead to corrections of order $\xi = g^2/\lambda$. From the imaginary part of the corresponding self-energy the  
fermion production rate then scales proportional to $\xi$. We have explicitly verified this from the numerics for small $\xi$ by varying $g$. As a consequence, the shown numerical results for $g = 0.1$ underestimate the even more dramatic production for the physically interesting situation with $\xi \gtrsim 1$. We emphasize that the observed phenomenon of dynamic enhancement of quantum corrections is not restricted to chaotic inflation and parametric resonance. The large correlations lead to universal behavior~\cite{Berges:2008wm}, and the discussion for tachyonic preheating or spinodal instabilities would follow along the same lines. 

In view of the dramatic differences between LO and NLO results, it remains to demonstrate that higher-order quantum corrections do not change our understanding. Since the interaction term (\ref{eq:interaction}) is quadratic in the fermion fields, we 'integrate out' the fermions on a lattice and study the remaining bosonic effective theory by numerical integration of the equations of motion and Monte Carlo sampling~\cite{Aarts:1998td}. This requires numerical integration of the fermion propagator, which becomes feasible in $3+1$ dimensions using "low-cost" fermion algorithms suggested recently in Ref.~\cite{Borsanyi:2008eu}. 
The approach treats the fermions exactly. However, the inflaton dynamics becomes classical-statistical, which has been shown in the past to be an accurate description for the sufficiently early times we are considering here~\cite{Berges:2002cz,Berges:2008wm}.

\begin{figure}[t]
\centering
\includegraphics[scale=0.7]{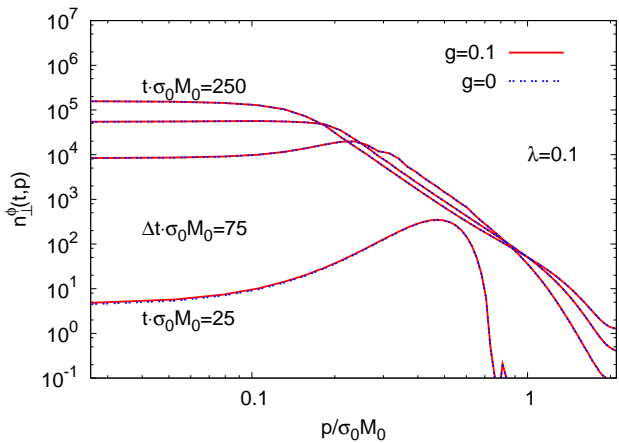}
 \caption{Inflaton occupation number of transverse modes.}
 \label{fig:ntrans}
\end{figure}

Remarkably, the lattice simulation results shown in Figure~\ref{fig:nferm} (symbols) agree well with the NLO approximation up to momenta of about $1.3/\sigma_0 M_0$~\footnote{For the $64^3$ lattice with spacing $a_s=1.5/\sigma_0 M_0$ the observed differences at somewhat larger $|{\bf p}|$ are found to depend on the spatial Wilson term, which we have to add to suppress doublers for the employed Wilson fermions. A possible better agreement on even larger lattices and a systematic study of cutoff dependencies also for those momenta is beyond the scope of the present work.} for the employed small $\xi$. We conclude that the distribution indeed exhibits a thermal value $n_{\rm eq}(0) = 1/2$ for massless fermions at low momenta, which drops at a characteristic scale $\sim \sigma_0 M_0$. We emphasize that the inflaton is still far from equilibrium at this stage. Figure~\ref{fig:ntrans} shows the inflaton occupation number as a function of momentum at different times, following standard conventions of Refs.~\cite{Traschen:1990sw,Khlebnikov:1996mc,Berges:2002cz}. In particular, at time $t = 250/\sigma_0 M_0$ the inflaton clearly approaches a nonthermal power-law distribution $\sim 1/|{\bf p}|^4$ characteristic for strong turbulence~\cite{Berges:2008wm}.

\begin{figure}[t]
 \centering
 \includegraphics[scale=0.7]{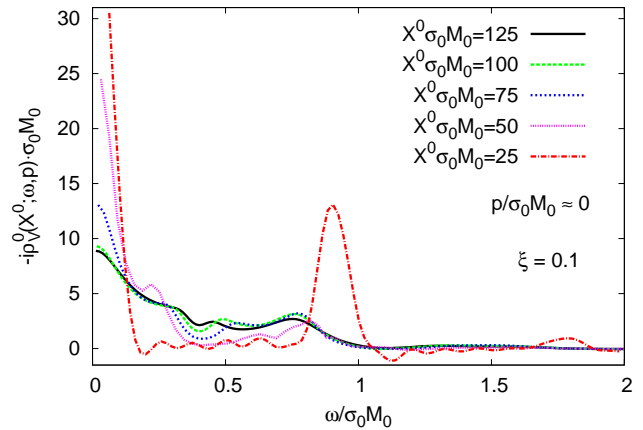}
\caption{Nonequilibrium fermion spectral function.}
\label{fig:fermi_spectral}
\end{figure}

A potentially important application of fermion preheating is the possibility to produce very heavy fermions, which may be interesting for the problems of dark matter and ultrahigh energy cosmic rays. This question can be addressed using the dynamical spectral function (\ref{eq:Fdef}). For the discussion of $\rho(x,y)$ we introduce Wigner coordinates, $X=(x+y)/2$ and $x-y$. For the spatially homogeneous system the spectral function then depends in addition only on time $X^0$. For the temporal component $\rho_V^0 = \mathrm{tr} ( \gamma^0 \rho )/4$ the sum over all frequencies is normalized to one,
$-i \int {\rm d} \omega/(2\pi)\, \rho^0_V(X^0;\omega,\mathbf{p}) = 1$,
which corresponds to the equal-time anticommutation relation for fermions. Figure~\ref{fig:fermi_spectral} shows NLO results at different times and $|{\bf p}|/\sigma_0 M_0\simeq 0$. Initially (dash-dotted curve) we find a double peak at $\omega \simeq 0$, corresponding to massless fermions, and $\omega \simeq g \phi_0/N_f$ characterizing very heavy fermion states. At later times these peaks become much broader and lower, which is characteristic for strong correlations without a quasiparticle interpretation. 

To conclude, we note that for any boson field with nonperturbatively large occupation number and Yukawa-type coupling to fermions a similar enhancement of quantum corrections may be observed. This concerns, in particular, also quantum chromodynamics in the presence of plasma instabilities. In the context of heavy-ion collisions for large gluon occupation numbers $\sim 1/\alpha_s$, where the running gauge coupling $\alpha_s$ is small at sufficiently high energies, a similar enhancement of quantum corrections would have dramatic consequences for quark production. 

We thank Denes Sexty for collaboration on related topics. This work is supported by the BMBF Grant No. 06DA9018.


\begin{thebibliography}{10}
  
%\cite{Traschen:1990sw}
\bibitem{Traschen:1990sw}
  J.~H.~Traschen and R.~H.~Brandenberger,
  %``PARTICLE PRODUCTION DURING OUT-OF-EQUILIBRIUM PHASE TRANSITIONS,''
  Phys.\ Rev.\  D {\bf 42} (1990) 2491;
  %%CITATION = PHRVA,D42,2491;%%  
%\cite{Kofman:1994rk}
%\bibitem{Kofman:1994rk}
  L.~Kofman, A.~D.~Linde and A.~A.~Starobinsky,
  %``Reheating after inflation,''
  Phys.\ Rev.\ Lett.\  {\bf 73} (1994) 3195.
  %[arXiv:hep-th/9405187].
  %%CITATION = PRLTA,73,3195;%%
  
%\cite{Khlebnikov:1996mc}
\bibitem{Khlebnikov:1996mc}
  S.~Y.~Khlebnikov and I.~I.~Tkachev,
  %``Classical decay of inflaton,''
  Phys.\ Rev.\ Lett.\  {\bf 77} (1996) 219.
  %[arXiv:hep-ph/9603378].
  %%CITATION = PRLTA,77,219;%%
  
%\cite{Berges:2002cz}
\bibitem{Berges:2002cz}
  J.~Berges and J.~Serreau,
  %``Parametric resonance in quantum field theory,''
  Phys.\ Rev.\ Lett.\  {\bf 91} (2003) 111601.
  %[arXiv:hep-ph/0208070].
  %%CITATION = PRLTA,91,111601;%%

\bibitem{RHIC} J.~Adams et.\ al., Nucl.\ Phys.\ A {\bf 757} (2005) 102; K.~Adcox et.\ al.,
{\it ibid.} 757 (2005) 184; I.~Arsene et.\ al., {\it ibid.} {\bf 757} (2005) 1; B.~B.~Back et.\ al., {\em ibid.} 757 (2005) 28.
  
%\cite{Baacke:1998di}
\bibitem{Baacke:1998di}
  J.~Baacke, K.~Heitmann and C.~Patzold,
  %``Nonequilibrium dynamics of fermions in a spatially homogeneous scalar
  %background field,''
  Phys.\ Rev.\  D {\bf 58} (1998) 125013;
  %[arXiv:hep-ph/9806205].
  %%CITATION = PHRVA,D58,125013;%%  
%\cite{Greene:1998nh}
%\bibitem{Greene:1998nh}
  P.~B.~Greene and L.~Kofman,
  %``Preheating of fermions,''
  Phys.\ Lett.\  B {\bf 448} (1999) 6;
  %[arXiv:hep-ph/9807339].
  %%CITATION = PHLTA,B448,6;%%
%\cite{Giudice:1999fb}
%\bibitem{Giudice:1999fb}
  G.~F.~Giudice {et al.},
  %, M.~Peloso, A.~Riotto and I.~Tkachev,
  %``Production of massive fermions at preheating and leptogenesis,''
  JHEP {\bf 9908} (1999) 014; 
  %[arXiv:hep-ph/9905242].
  %%CITATION = JHEPA,9908,014;%%
%\cite{GarciaBellido:2000dc}
%\bibitem{GarciaBellido:2000dc}
  J.~Garcia-Bellido, S.~Mollerach and E.~Roulet,
  %``Fermion production during preheating after hybrid inflation,''
  JHEP {\bf 0002} (2000) 034; 
  %[arXiv:hep-ph/0002076].
  %%CITATION = JHEPA,0002,034;%%
%\cite{Peloso:2000hy}
%\bibitem{Peloso:2000hy}
  M.~Peloso and L.~Sorbo,
  %``Preheating of massive fermions after inflation: Analytical results,''
  JHEP {\bf 0005} (2000) 016.
  %[arXiv:hep-ph/0003045].
  %%CITATION = JHEPA,0005,016;%%  
%\cite{Baacke:2007ca}
%\bibitem{Baacke:2007ca}
  J.~Baacke, N.~Kevlishvili and J.~Pruschke,
  %``Quantum back-reaction of the superpartners in a large-N supersymmetric
  %hybrid model,''
  JCAP {\bf 0706} (2007) 004. 
  %[arXiv:hep-th/0702009].
  %%CITATION = JCAPA,0706,004;%%

%\cite{Gelis:2005pb}
\bibitem{Gelis:2005pb}
  F.~Gelis, K.~Kajantie and T.~Lappi,
  %``Chemical thermalization in relativistic heavy ion collisions,''
  Phys.\ Rev.\ Lett.\  {\bf 96} (2006) 032304.
  %[arXiv:hep-ph/0508229].
  %%CITATION = PRLTA,96,032304;%%

%\cite{Berges:2002wr}
\bibitem{Berges:2002wr}
  J.~Berges, S.~Borsanyi and J.~Serreau,
  %``Thermalization of fermionic quantum fields,''
  Nucl.\ Phys.\  B {\bf 660} (2003) 51.
  %[arXiv:hep-ph/0212404].
  %%CITATION = NUPHA,B660,51;%%     

%\cite{Berges:2009bx}
\bibitem{Berges:2009bx}
  J.~Berges, J.~Pruschke and A.~Rothkopf,
  %``Instability-induced fermion production in quantum field theory,''
  Phys.\ Rev.\  D {\bf 80} (2009) 023522.
  %[arXiv:0904.3073 [hep-ph]].
  %%CITATION = PHRVA,D80,023522;%%

%\cite{Aarts:1998td}
\bibitem{Aarts:1998td}
  G.~Aarts and J.~Smit,
  %``Real-time dynamics with fermions on a lattice,''
  Nucl.\ Phys.\  B {\bf 555} (1999) 355.
  %[arXiv:hep-ph/9812413].
  %%CITATION = NUPHA,B555,355;%%
  
%\cite{Borsanyi:2008eu}
\bibitem{Borsanyi:2008eu}
  S.~Borsanyi and M.~Hindmarsh,
  %``Low-cost fermions in classical field simulations,''
  Phys.\ Rev.\  D {\bf 79} (2009) 065010.
  %[arXiv:0809.4711 [hep-ph]].
  %%CITATION = PHRVA,D79,065010;%%

%\cite{Berges:2008wm}
\bibitem{Berges:2008wm}
  J.~Berges, A.~Rothkopf and J.~Schmidt,
  %``Non-thermal fixed points: effective weak-coupling for strongly correlated
  %systems far from equilibrium,''
  Phys.\ Rev.\ Lett.\  {\bf 101} (2008) 041603.
  %[arXiv:0803.0131 [hep-ph]].
  %%CITATION = PRLTA,101,041603;%%
    
%\cite{Micha:2002ey}
\bibitem{Micha:2002ey}
  R.~Micha and I.~I.~Tkachev,
  %``Relativistic turbulence: A long way from preheating to equilibrium,''
  Phys.\ Rev.\ Lett.\  {\bf 90} (2003) 121301.
  %[arXiv:hep-ph/0210202].
  %%CITATION = PRLTA,90,121301;%%  
    
%\cite{Berges:2001fi}
\bibitem{Berges:2001fi}
  J.~Berges,
  %``Controlled nonperturbative dynamics of quantum fields out of
  %equilibrium,''
  Nucl.\ Phys.\  A {\bf 699} (2002) 847.
  %[arXiv:hep-ph/0105311].
  %%CITATION = NUPHA,A699,847;%%
  %\cite{Aarts:2002dj}
%\bibitem{Aarts:2002dj}
  G.~Aarts {\it et al.}, 
  %D.~Ahrensmeier, R.~Baier, J.~Berges and J.~Serreau,
  %``Far-from-equilibrium dynamics with broken symmetries from the 2PI-1/N
  %expansion,''
  Phys.\ Rev.\  D {\bf 66} (2002) 045008.
  %[arXiv:hep-ph/0201308].
  %%CITATION = PHRVA,D66,045008;%%  

%\cite{Berges:2008pc}
\bibitem{Berges:2008pc}
  J.~Berges,
  %``What the inflaton might tell us about RHIC/LHC,''
  Nucl.\ Phys.\  A {\bf 820} (2009) 65C.
  %[arXiv:0811.4401 [hep-ph]].
  %%CITATION = NUPHA,A820,65C;%%                 
\end{thebibliography}
\end{document}